\documentclass{KapProc} 
\usepackage{astrobib}
\usepackage{psfig}
\usepackage{astro}

\let\footnote\savefootnote

\let\footnotetext\savefootnotetext 

\let\footnoterule\savefootnoterule 

\kluwerbib

\bibstyle{aa}
\bibdata{aamnemonic,../../Review/review}
\begin{document}
\articletitle{Host galaxies and nuclear\\structure of AGN with \\H$_2$O megamasers as seen with HST$^1$}
\chaptitlerunninghead{H$_2$O megamasers seen with HST}

\author{Heino Falcke, Christian Henkel}
\affil{Max-Planck-Institut f\"ur Radioastronomie, Auf dem H\"ugel 69,
53121 Bonn, Germany}
\email{hfalcke@mpifr-bonn.mpg.de, chenkel@mpifr-bonn.mpg.de}

\author{Andrew S. Wilson\footnote{Adjunct Astronomer, Space Telescope Science Institute}}
\affil{Astronomy Department, University of Maryland, College Park,
MD 20742-2421, USA}
\email{wilson@astro.umd.edu}

\author{James A. Braatz}
\affil{National Radio Astronomy Observatory, P.O. Box 2,
Green Bank, WV 24944, USA}
\email{jbraatz@nrao.edu}

\begin{abstract}\footnotetext[1]{Paper to appear in: ``QSO hosts and Their Environments'', Granada, January
2001, Eds. I. M\'arquez P\'erez et al., Kluwer Academic Press}
We present results of an HST survey in H$\alpha$ and continuum filters
of a sample of H$_2$O megamaser galaxies compiled by Braatz et al.,
all of which contain AGN. These observations allow us to study the
AGN/host-galaxy connection, e.g. study the relation between the parsec
scale masing disk/torus, bipolar outflows, and large scale properties
of the galaxies such as dust lanes, signs for interaction, and galaxy
types. A number of galaxies indeed show large-scale bi-polar H$\alpha$
structures which, however, are more reminiscent of outflows then
excitation cones. Most megamaser galaxies are found in spiral
galaxies. Only one galaxy in the original sample is known to be an
elliptical and one galaxy imaged by us shows clear signs of
interactions. In all cases we see evidence for obscuration of the
nucleus (e.g. dust-lanes) in our color maps and the disk galaxies are
preferentially edge on. This suggests that the nuclear masing disk has
some relation to the large scale properties of the galaxy and the dust
distribution.
\end{abstract}


\section{Introduction} 
After the discovery of extragalactic water
vapor masers in star-forming regions
\cite{ChurchwellWitzelPauliny-Toth1977} a search started for other
possible extragalactic H$_2$O masers. In the late seventies and early
eighties, this work lead to to the discovery of five so-called
megamasers in the nuclei of NGC 4945, the Circinus galaxy, NGC 1068,
NGC 4258, and NGC 3079 which are $10^6$ times more powerful than the
commonly observed galactic masers. Only much later, after an intense
search of a distance limited galaxy sample, Braatz, Wilson, \& Henkel
(1994, 1996) were able to increase the number of known megamasers by a
factor of three, thus making for the first time a statistical analysis
of this rare phenomenon possible.
\nocite{BraatzWilsonHenkel1994,BraatzWilsonHenkel1996}

It is obvious that the H$_2$O megamaser phenomenon must have something
to do with nuclear activity as all such megamaser sources are either
in Seyfert or Liner galaxies and the emission is always concentrated
at the nucleus. Interestingly, Seyferts of type 1 are absent from this
group. The standard interpretation for the difference between Sy 1 and
Sy 2 is that a torus surrounds the nucleus which obscures the central
engine (black hole and accretion disk) for large inclination angles.

It therefore appears reasonable to infer that the masers trace
molecular material associated with this torus or an accretion disk
that feeds the nucleus. This, and the small number of H$_2$O
megamasers known, suggests that the maser emission occurs for certain,
very restricted, viewing angles, perhaps where the line of sight is
along the plane of the molecular disk or torus.

This notion was recently confirmed in great detail by VLBI
observations of the megamaser in NGC 4258
\cite{MiyoshiMoranHerrnstein1995,GreenhillJiangMoran1995}.  The
positions and velocities of the central, the red- and the blue-shifted
H$_2$O maser lines show that the masing region is in a thin disk
structure, rotating on Keplerian orbits around a central mass of
$3.6\cdot10^7M_\odot$ at a distance of 0.13 pc from the nucleus. This
observations proved for the first time the existence of a small scale
molecular disk surrounding an active nucleus. Recently, VLBI maps have
been obtained for even more maser sources: while NGC 4945
\cite{GreenhillMoranHerrnstein1997} and NGC 3079
\cite{TrotterGreenhillMoran1998} also show a structure interpreted as
a disk, the masers in NGC1068 show weak emission from a region
presumably shocked by the nuclear jet at a projected galactocentric
distance of $\sim$30 pc from the nucleus
\cite{GallimoreBaumHenkel2001} in addition to the masers in the disk
\cite{GreenhillGwinnAntonucci1996}.

Although plausible scenarios for the megamaser phenomenon exist
(e.g.,~\citeNP{NeufeldMaloney1995}) it is by no means clear how the
obscuring torus and the masing disk are related. The masing disk may
be the innermost part of a molecular accretion disk adjacent to the
torus or just be the thin, central plane of the torus in which the
column density is high enough for strong amplification. Alternatively,
it might also be dynamically independent from ``the torus'' with a
strong misalignment between their axes. In the most simple-minded
picture, however, one would hope to find that masing disk, obscuring
torus and the extended molecular cloud distribution form one coherent
accretion structure feeding the central engine.

We have therefore investigated a sample of megamaser host galaxies
with the Hubble-Space-Telescope (HST) to look for large scale evidence
of the molecular material, some of which we describe in the following.

\section{Observations} The sample was selected from the list of 16
known megamasers known in 1996 and published by
\citeN{BraatzWilsonHenkel1996}. We observed all galaxies that had not
been previously observed with HST, which included: NGC449 (Mrk 1),
NGC1386, NGC2639, NGC3079, IC2560, NGC4945, IRAS18333-6528,
TXS~2226-184, and IC1481.  Observations were made with WFPC2 to cover
a red and a green continuum filter, as well as the
H$\alpha$+[\ion{N}{2}]$\lambda\lambda$6548,6583 emission lines. The
filters used were F814W (red continuum), F547M (green continuum),
F673N and the Linear-Ramp-Filters (LRF) for galaxies at higher
redshifts. The pixel scales were either 0\farcs1/pixel or
0\farcs0455/pixel depending on whether LRF usage required imaging on
the Wide Field or Planetary Camera. All observations were made within
one orbit.

\begin{figure}[ht]
\centerline{\hbox{\psfig{figure=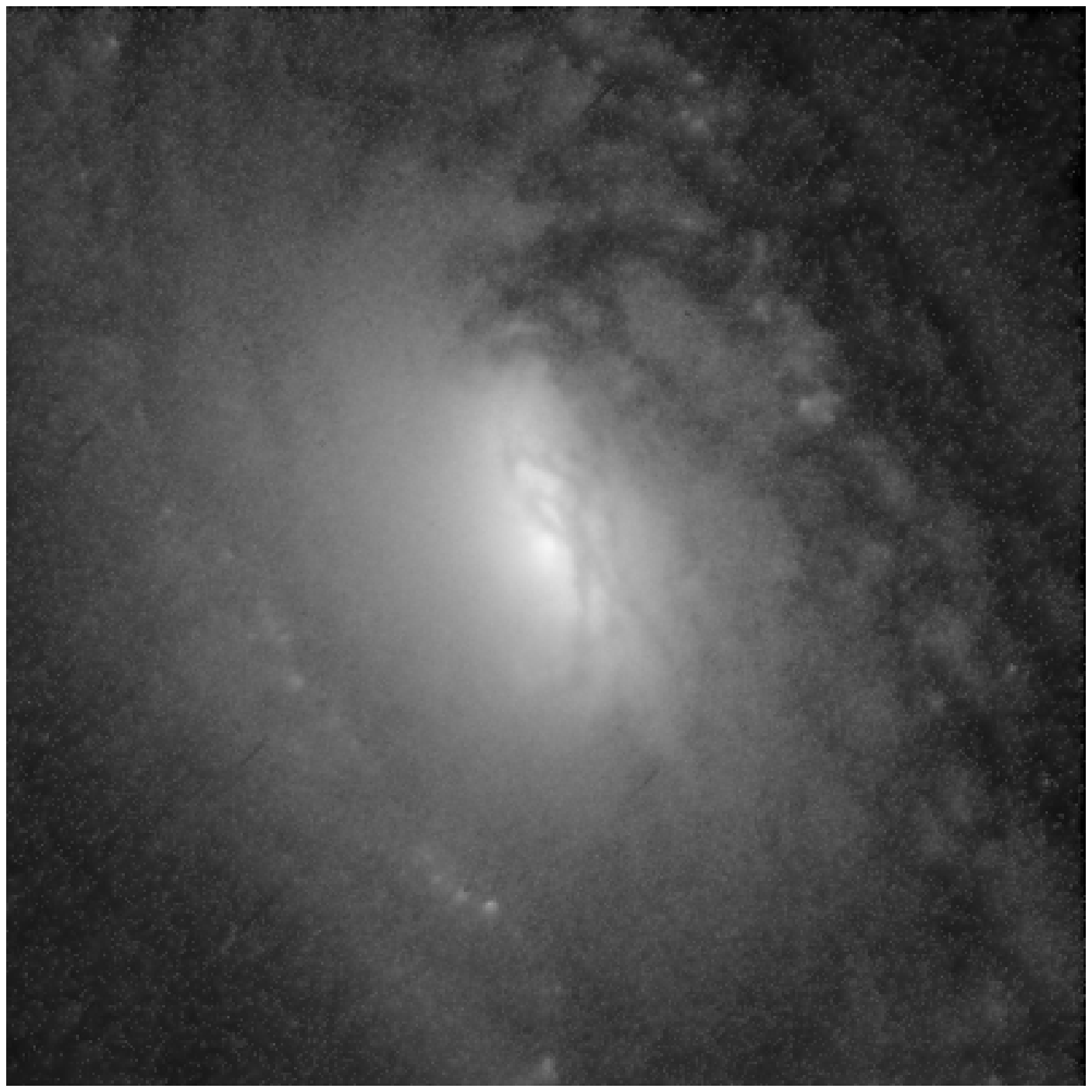,height=0.3\textheight}\hspace{0.5cm}\psfig{figure=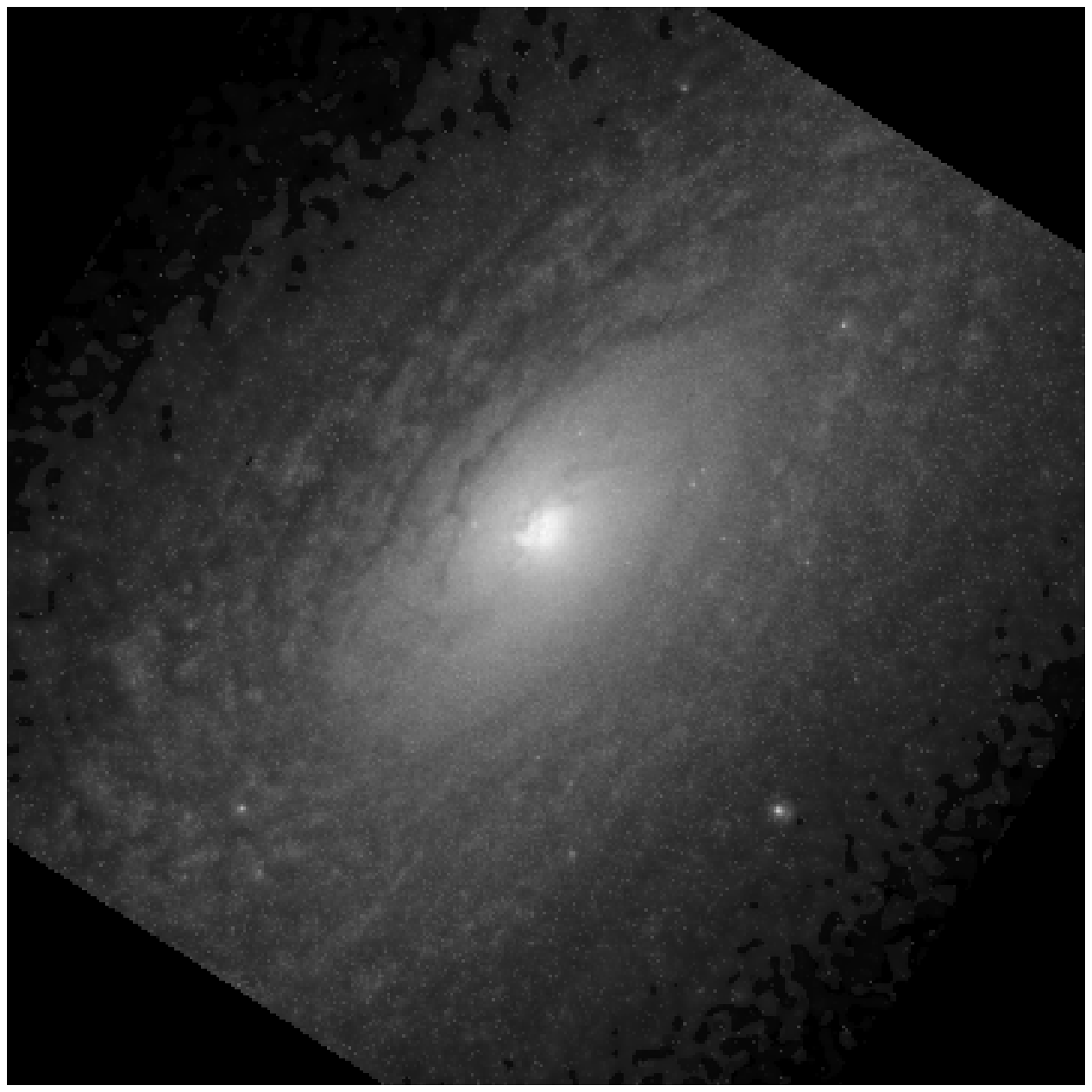,height=0.3\textheight}}}
\caption[]{Gray-scale representation of three-color HST images of the
megamaser galaxies NGC1386 (left) and NGC2639 (right). North is up and 
East to the left. The images are 18\farcs25 and 32\farcs35 on each
side respectively.}
\end{figure}

The images were processed through the standard Wide-Field and
Planetary Camera 2 (WFPC2) pipeline data reduction at the Space
Telescope Science Institute. Further data reduction was done in IRAF
and included: cosmic ray rejection, flux calibration, and rotation to
the cardinal orientation.  The galaxy continuum near the
H$\alpha$+[\ion{N}{2}] line was determined by combining the red and
green continuum images, scaled to the filter width of the narrow-band
filter and weighted by the relative offset of their mean wavelengths
from the redshifted H$\alpha$+[\ion{N}{2}] emission. From the two
broad-band images and the emission-line image, we constructed
three-color (RGB) maps, where, for better contrast, the
H$\alpha$+[\ion{N}{2}] image was assigned the green channel, and the
F814W and F547M filter images were assigned the red and blue channels
respectively. Galaxy profile fits were made to the red continuum
images. More details and first results for one galaxy (TXS~2226--184)
are given in
\citeN{FalckeWilsonHenkel2000}.

\section{Results}

\begin{figure}[ht]
\centerline{\hbox{\psfig{figure=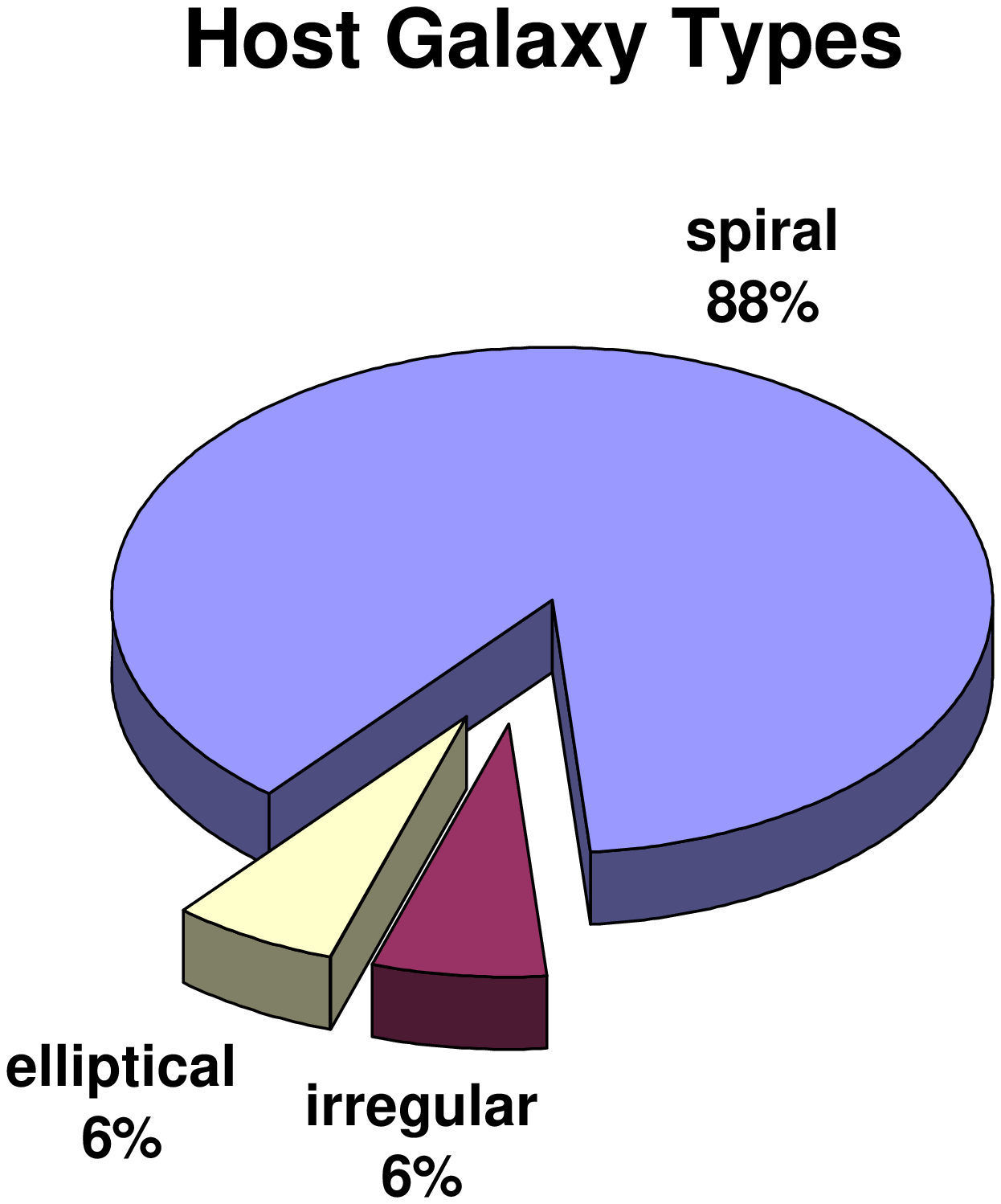,width=0.3\textheight,bblly=3.7cm,bbllx=3.5cm,bburx=17cm,bbury=16.5cm,clip=}\hspace{0.5cm}\psfig{figure=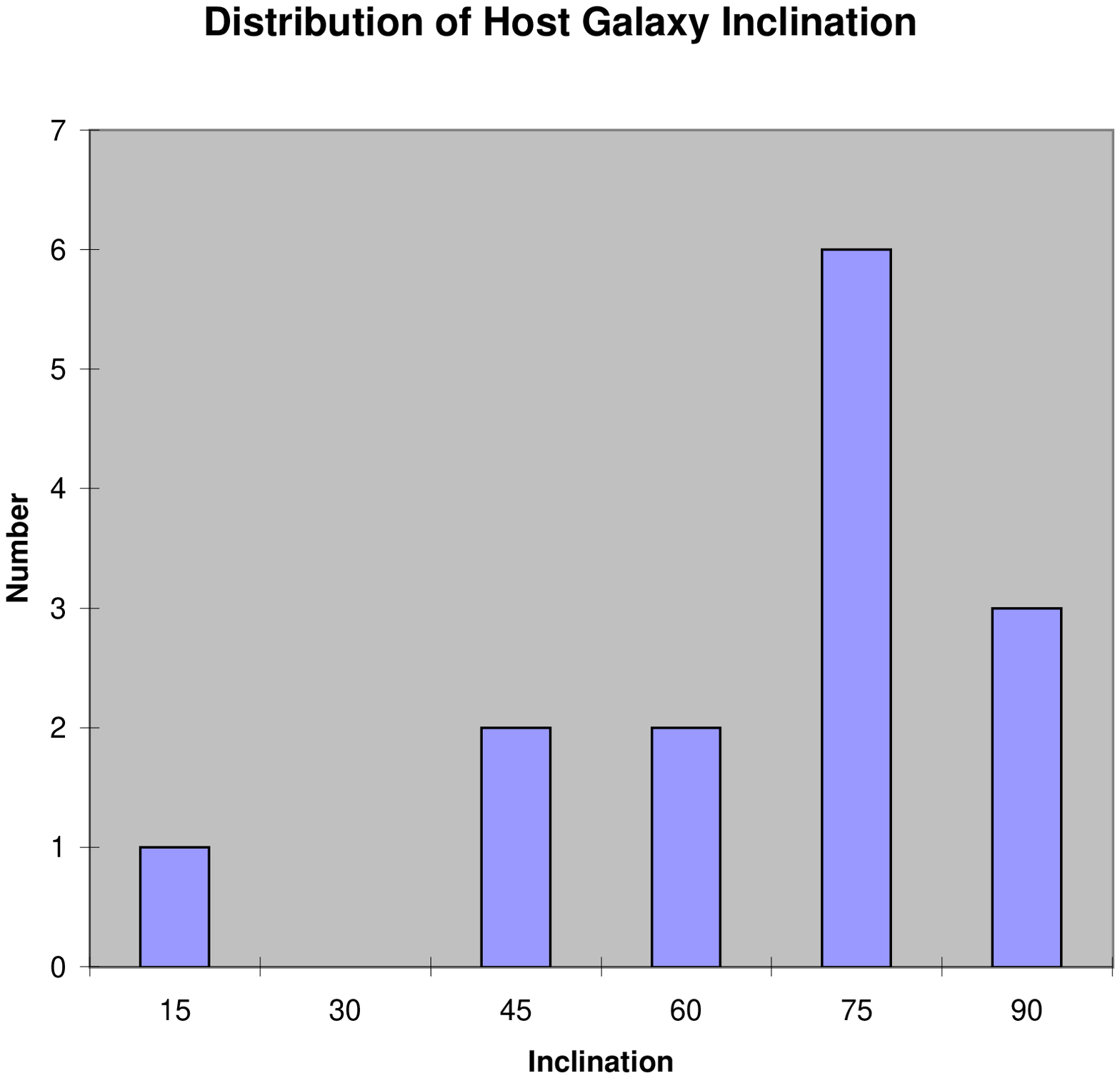,width=0.3\textheight,bblly=3.2cm,bbllx=2.4cm,bburx=18.9cm,bbury=18.6cm}}}
\caption[]{Left: The distribution of host galaxy types for all
megamasers in the Braatz et al.~(1996) sample determined from our HST
observations and the literature. Right: Distribution of host galaxy
inclination. The majority of megamaser galaxies are edge-on spirals.}
\end{figure}

The HST images contain a wealth of information. In many cases dust
lanes and spiral arms are clearly visible and in basically all
galaxies we resolve the narrow-line region (NLR). The most spectacular
structures are the well-known H$\alpha$-bubble in NGC3079 (see also
\citeNP{CecilBland-HawthornVeilleux2001}), a fan-shape region in
NGC4945 \cite{MarconiOlivavanderWerf2000}, and linear jet-like
features in TXS~2226-184 \cite{FalckeWilsonHenkel2000} and NGC~1386
\cite{FerruitWilsonMulchaey2000}. In these cases the structure might
be mainly related to an outflow rather than an excitation cone. The
water vapor maser emission in these galaxies has been imaged with VLBI
 (either published or private communication: L. Greenhill and
J. Braatz). In those four cases the presumed axis of the nuclear disk
(even though the disk interpretation is not secure in all cases) seems
to be perpendicular within less than 30$^\circ$ to the closest nuclear
dust lane, as would be the case if the maser disks and the dust lanes
tend to align. On the other hand the large scale radio or H$\alpha$
axis is not always perpendicular to the disk axis (see, e.g.,~NGC3079,
\citeNP{TrotterGreenhillMoran1998}).

In all galaxies of our sample, we see clear evidence for the presence
of large scale dust obscuration. In most, i.e. seven, cases we
directly see relatively well-defined narrow dust-lanes passing through
the nucleus; in the remaining two cases we see an elongated nucleus
and reddening on one side of the inner galaxy.

The profile fitting of the continuum images yielded a significant
preference for disk models over bulge-only/elliptical models. We also
see clear signs of spiral arms in a number of the disk galaxies. Only
for one galaxy, IC1481, do we obtain inconclusive results. The optical
appearance is very irregular suggesting that this might be the site of
an ongoing galaxy merger. One galaxy, TXS~2226-184, was re-classified
as a spiral (see \citeNP{FalckeWilsonHenkel2000} -- it was originally
thought to be an elliptical). If we supplement our sample with
literature data we find that only one out of 16 galaxies is an
elliptical and one is irregular (see Fig.~2a) -- the vast majority of
megamaser host galaxies are indeed spirals.

From the observed ellipticity we can derive inclinations for the
spiral galaxies in our sample. This shows that the majority of the
galaxies are highly inclined with only a few face-on odd-balls (Fig.~2b).

\section{Conclusions}
\citeN{BraatzWilsonHenkel1997} suggested that the nuclear
properties of AGN with megamasers might be connected with their large
scale structure. Our investigation seems to confirm this
trend. Megamaser galaxies seem to be preferentially edge-on and in
spiral galaxies. After TXS~2228-184 was reclassified there is only one
elliptical galaxy left (NGC~1052), which, however, has masers along
the jet \cite{ClaussenDiamondBraatz1998} and hence may be different
from the majority of galaxies which have been interpreted in terms of
masers from molecular disks\footnote[2]{The same may be true for the
recently discovered megamaser in Mrk~348 \cite{FalckeHenkelPeck2000},
the galactic disk of which is very much face-on. VLBA observations
of this galaxy show the masers to
be at the tip of an expanding jet (Peck et al., in preparation).}.

The fact that we always see some obscuring material in the HST images
directly on scales of several tens to hundreds of parsecs could
indicate that the masing structure (the `torus') on the
milli-arcsecond (i.e., sub-pc) scale is related to large scale dust
lanes and is thus not an isolated nuclear feature.  Confirmation of
this hypothesis may come from further VLBA-observations which could
reveal the orientation and geometry of the nuclear disk in relation to
the dust lanes seen with HST.

\begin{acknowledgments}
This work was supported by DFG grant 358/1-1 \& 2, NASA grants
NAG8-1027 and HST GO 7278, and NATO grant SA.5-2-05 (GRG 960086)318/96.
\end{acknowledgments}


\section*{References}
\bibitem[\protect\citeauthoryear{{Braatz}, {Wilson}, \& {Henkel}}{{Braatz}
  et~al.}{1994}]{BraatzWilsonHenkel1994}
{Braatz} J.~A., {Wilson} A.~S.,  {Henkel} C., 1994, \apjl, 437, L99

\bibitem[\protect\citeauthoryear{{Braatz}, {Wilson}, \& {Henkel}}{{Braatz}
  et~al.}{1996}]{BraatzWilsonHenkel1996}
{Braatz} J.~A., {Wilson} A.~S.,  {Henkel} C., 1996, \apjs, 106, 51

\bibitem[\protect\citeauthoryear{{Braatz}, {Wilson}, \& {Henkel}}{{Braatz}
  et~al.}{1997}]{BraatzWilsonHenkel1997}
{Braatz} J.~A., {Wilson} A.~S.,  {Henkel} C., 1997, \apjs, 110, 321

\bibitem[\protect\citeauthoryear{{Cecil} et~al.}{{Cecil}
  et~al.}{2001}]{CecilBland-HawthornVeilleux2001}
{Cecil} G., {Bland-Hawthorn} J., {Veilleux} S.,  {Filippenko} A.~V., 2001,
  \apj, in press

\bibitem[\protect\citeauthoryear{{Churchwell} et~al.}{{Churchwell}
  et~al.}{1977}]{ChurchwellWitzelPauliny-Toth1977}
{Churchwell} E., {Witzel} A., {Pauliny-Toth} I., et~al., 1977, \aap, 54, 969

\bibitem[\protect\citeauthoryear{{Claussen} et~al.}{{Claussen}
  et~al.}{1998}]{ClaussenDiamondBraatz1998}
{Claussen} M.~J., et al., 1998, \apjl, 500, L129

\bibitem[\protect\citeauthoryear{{Falcke} et~al.}{{Falcke}
  et~al.}{2000a}]{FalckeHenkelPeck2000}
{Falcke} H., {Henkel} C., {Peck} A.~B., et~al., 2000a, \aap, 358, L17

\bibitem[\protect\citeauthoryear{{Falcke} et~al.}{{Falcke}
  et~al.}{2000b}]{FalckeWilsonHenkel2000}
{Falcke} H., {Wilson} A.~S., {Henkel} C., {Brunthaler} A.,  {Braatz} J.~A.,
  2000b, \apjl, 530, L13

\bibitem[\protect\citeauthoryear{{Ferruit}, {Wilson}, \& {Mulchaey}}{{Ferruit}
  et~al.}{2000}]{FerruitWilsonMulchaey2000}
{Ferruit} P., {Wilson} A.~S.,  {Mulchaey} J., 2000, \apjs, 128, 139

\bibitem[\protect\citeauthoryear{{Gallimore} et~al.}{{Gallimore}
  et~al.}{2001}]{GallimoreBaumHenkel2001}
{Gallimore} J., {Baum} S., {Henkel} C.,  {et al.} , 2001, \apj, submitted

\bibitem[\protect\citeauthoryear{{Greenhill} et~al.}{{Greenhill}
  et~al.}{1996}]{GreenhillGwinnAntonucci1996}
{Greenhill} L.~J., {Gwinn} C.~R., {Antonucci} R.,  {Barvainis} R., 1996, \apjl,
  472, L21

\bibitem[\protect\citeauthoryear{{Greenhill} et~al.}{{Greenhill}
  et~al.}{1995}]{GreenhillJiangMoran1995}
{Greenhill} L.~J., {Jiang} D.~R., {Moran} J.~M., et~al., 1995, \apj, 440, 619

\bibitem[\protect\citeauthoryear{{Greenhill}, {Moran}, \&
  {Herrnstein}}{{Greenhill} et~al.}{1997}]{GreenhillMoranHerrnstein1997}
{Greenhill} L.~J., {Moran} J.~M.,  {Herrnstein} J.~R., 1997, \apjl, 481, L23

\bibitem[\protect\citeauthoryear{{Marconi} et~al.}{{Marconi}
  et~al.}{2000}]{MarconiOlivavanderWerf2000}
{Marconi} A., {Oliva} E., {van der Werf} P.~P., et~al., 2000, \aap, 357, 24

\bibitem[\protect\citeauthoryear{{Miyoshi} et~al.}{{Miyoshi}
  et~al.}{1995}]{MiyoshiMoranHerrnstein1995}
{Miyoshi} M., {Moran} J., {Herrnstein} J., et~al., 1995, \nat, 373, 127

\bibitem[\protect\citeauthoryear{{Neufeld} \& {Maloney}}{{Neufeld} \&
  {Maloney}}{1995}]{NeufeldMaloney1995}
{Neufeld} D.~A.,  {Maloney} P.~R., 1995, \apjl, 447, L17

\bibitem[\protect\citeauthoryear{{Trotter} et~al.}{{Trotter}
  et~al.}{1998}]{TrotterGreenhillMoran1998}
{Trotter} A.~S., et~al., 1998, \apj, 495, 740


\end{document}